# Phonon transition across an isotopic interface

Ning Li[1,2,3#], Ruochen Shi[1,2#], Yifei Li[4#], Ruishi Qi[1,2], Zhetong Liu[1,2,3], Fachen Liu[1,2,3], Yuehui Li[1,2], Xiaowen Zhang[1,2], Xiangdong Guo[5], Kaihui Liu[6,7], Ying Jiang[1,7,8], Xin-Zheng Li[6,8,9], Ji Chen[6,7,8]*, Lei Liu[4,8]*, En-Ge Wang[1,8,10,11]*, and Peng Gao[1,2,7,8]*

[1]International Center for Quantum Materials, School of Physics, Peking University, Beijing 100871, China.

[2]Electron Microscopy Laboratory, School of Physics, Peking University, Beijing 100871, China.

[3]Academy for Advanced Interdisciplinary Studies, Peking University, Beijing 100871, China.

[4]School of Materials Science and Engineering, Peking University, Beijing 100871, China.

[5]CAS Key Laboratory of Nanophotonic Materials and Devices, CAS Key Laboratory of Standardization and Measurement for Nanotechnology, CAS Center for Excellence in Nanoscience, National Center for Nanoscience and Technology, Beijing 100190, China.

[6]Institute of Condensed Matter and Material Physics, Frontiers Science Center for Nano-optoelectronics, School of Physics, Peking University, Beijing 100871, China.

[7]Collaborative Innovation Center of Quantum Matter, Beijing 100871, China.

[8]Interdisciplinary Institute of Light-Element Quantum Materials and Research Center for Light-Element Advanced Materials, Peking University, Beijing 100871, China.

[9]State Key Laboratory for Artificial Microstructure and Mesoscopic Physics, Peking University, Beijing 100871, China.

[10]Songshan Lake Materials Laboratory, Dongguan 523808, China.

[11]School of Physics, Liaoning University, Shenyang 110036, China.

[#]*These authors contributed equally: Ning Li, Ruochen Shi, Yifei Li*

*Corresponding author. Email: ji.chen@pku.edu.cn, l_liu@pku.edu.cn, egwang@pku.edu.cn, p-gao@pku.edu.cn*

**Abstract**：

Natural materials usually consist of isotopic mixtures, for which different isotopic ratios can lead to distinct material properties such as thermal conductivity and nucleation process [1, 2, 3, 4, 5, 6, 7, 8, 9]. However, the knowledge of isotopic interface remains largely unexplored mainly due to the challenges in isotopic identification and property measurement at an atomic scale [10, 11]. Here, by using monochromated electron energy-loss spectroscopy in a scanning transmission electron microscope, we reveal momentum-transfer-dependent lattice vibration behavior at an artificial h-$^{10}$BN/h-$^{11}$BN heterostructure with sub-unit-cell resolution. We find the vibrational energy changes across the isotopic interface gradually, featuring a wide transition regime, which suggests strong delocalization of the out-of-plane optical phonons at the interface. In addition, we identify phonons near the Brillouin zone center have a transition regime ~3.34 nm (10 atomic layers), whereas phonons at the Brillouin zone boundary transition in ~1.66 nm (5 atomic layers). We propose that the isotope-induced charge effect at the interface accounts for the distinct delocalization behavior. Moreover, intra-atomic layer variation of vibration energy is also sensitive to the momentum transfer, thus at the interface it depends on both of momentum transfer and mass change. The revealed lattice vibration behavior at an isotopic interface provides new insights to understand the isotopic effects on properties in natural materials.

## Introduction

Isotopes of an element have the same number of protons, but different in numbers of neutrons and hence atomic mass [1], leading to different nuclei-governed properties, such as thermal conductivity [2, 3, 4, 6, 12], elasticity [7], and nuclear reactions [8, 9]. The isotopic labeling is widely used as tracers in molecular labeling [13, 14, 15], chemical reactions [16, 17, 18], and radiometric dating [19, 20, 21] due to the similar chemical properties. However, recently exotic phenomena relating to the distinct chemical/electronic properties of the isotope-enriched materials have also been reported, i.e., superconducting transition temperature $T_c$ between $H_3S/D_3S$ and $LaH_{10}/LaD_{10}$ is found to be different under high compression [22, 23], the transmissivity of hydrogen isotopic water passing through graphene oxide membrane is very different [24], and the optical band gap of h-$^{11}$BN and h-$^{10}$BN strongly depends on the isotopic composition [25]. These phenomena open a new door for function design via isotope engineering, and also lead to essential questions for the naturally existing materials that commonly consist of isotopic mixture: are there new properties emerging at the isotope interface? For example, the isotope enriched materials have much enhanced thermal conductivity compared to natural ones [12, 26, 27], for which a well-accepted assumption is that the isotope mass disorder in natural materials disrupts the phonon transport accounting for the substantial reduction of thermal conductivity. However, in the case of heterogeneous distribution of isotopes, the presence of isotope interfaces in a natural material makes it similar to the "superlattice" case. Under such a circumstance, the heat transport behavior is intuitively different from completely random ones [28]. Nevertheless, such knowledge on the isotope interface and possible effects has been rarely discussed before and thus largely unknown mainly due to the challenges in isotopic identification and property measurement at the atomic scale for their interfaces.

The commonly-used isotope analysis methods are based on vibration detection, such as Raman [2, 3] and Infrared spectroscopy [29] with an energy resolution at the order of 1 cm$^{-1}$ to distinguish the phonons of isotopes, but they usually have a limited spatial resolution. Although the spatial resolution can be substantially improved by the tip-enhanced Raman spectroscopy [10] and scanning near-field optical microscopy [11], optical characterizations still lack the ability to resolve the atomic structure, and access to high-momentum phonons at the Brillouin zone (BZ) boundary due to the tiny momentum of photons. Recent advances in electron energy loss spectroscopy (EELS) in scanning transmission electron microscope (STEM) enable momentum-resolved vibrational measurements at nanometer/atomic scale [30, 31, 32, 33, 34, 35], providing new opportunities to detect the isotopes distribution [36], e.g., H/D -O bonds [37], $^{13}C/^{12}C$ -O bonds [38], and $^{13}C$-$^{13}C$ / $^{12}C$-$^{12}C$ bonds [39].

Here, we fabricate atomically sharp h-$^{10}$BN/h-$^{11}$BN interface and characterize the lattice vibration properties at the interface with sub-unit-cell spatial resolution and different momentum transfer. We find that the out-of-plane optical phonon modes (dubbed as ZO) at the interface are not atomically sharp. Instead, they are delocalized, featured by a gradual transition across the interface. In addition, the delocalization depends on the momentum transfer of the ZO mode, i.e., modes with small momentum transfer ($ZO_{low\ q}$ at BZ center) are delocalized ~3.34 nm, while the ones with large momentum transfer ($ZO_{high\ q}$ at BZ boundary) are delocalized ~1.66 nm. The different vibration amplitude of isotopes causes the different vibration dipole magnitude and thus momentum-dependent charge density at the interface, which is proposed to account for the different delocalization behavior. These two ZO modes also have different intra-atomic layer variation, i.e., the $ZO_{high\ q}$ is less sensitive to position variation (associated with the change of momentum transfer) as the phonon dispersion is flat at the BZ boundary. Furthermore, across the interface from the h-$^{10}$BN layer to h-$^{11}$BN layer, the vibrational energy of these two ZO phonons goes up first, then followed by a rapid drop. Such a complicated behavior can be qualitatively understood by the combination of changes in the scattering cross section and mass across the interface. The delocalization of phonon modes at an isotopic interface and the momentum-dependent behavior may affect the local thermal processes. These findings provide us with a new angle to understand the isotopic effects in natural materials and insights into tailoring property via isotopic engineering.

**Results**

We grow large h-$^{11}$BN and h-$^{10}$BN isotope crystals (supplemental figure 1) and then transfer a ~20-nm-thick h-$^{11}$BN and a ~100-nm-thick h-$^{10}$BN to SiO$_2$/Si substrate (methods) forming an atomically sharp h-$^{10}$BN/h-$^{11}$BN/a-SiO$_2$ (a-SiO$_2$: amorphous SiO$_2$) heterostructure [40]. The schematic of the assembled stack and HAADF image at the cross-section direction are shown in Fig. 1(a, b). First-principles calculations based on density functional perturbation theory (DFPT) of the phonon dispersion and total phonon density of states (DOS) for h-$^{10}$BN and h-$^{11}$BN (supplemental figure 2 and note 1) give an energy shift of ~5 meV for the in-plane phonon peaks and ~4 meV for the out-of-plane phonon peaks between these two isotopes, which should be sufficiently large to be captured by the STEM-EELS [38, 41].

Fig. 1c shows the vibrational EELS acquired across the heterostructure (orange rectangle in Fig. 1b). The vibrational signals from the in-plane modes of h-BN (160-200 meV), a-SiO$_2$ (125-155 meV, 100 meV), and out-of-plane modes of h-BN (80-100 meV) are outlined by green, white and blue dashed rectangles, separately. Note that the phonon polaritons of h-BN (~180 meV) [42] and a-SiO$_2$ (~130 meV) are also observable. For the analysis below, we carefully exclude the phonon polariton signals and focus on the h-BN phonons (see details in supplemental figure 3 and Note 2). Fig. 1d shows typical vibrational spectra at h-$^{10}$BN, h-$^{11}$BN regions and interface, denoted by the purple, orange, and green solid lines, respectively. For h-$^{10}$BN, there is a blue shift of ~6.1 meV for the in-plane band and ~3.3 meV for the out-of-plane band compared to that of h-$^{11}$BN. The interfacial spectrum basically lies between the two bulk spectra. EELS mappings with different energy windows (marked by the colored stripes in Fig. 1d) are displayed in Fig. 1e, corresponding to the acquisition region denoted by the white box region in Fig. 1b. The interface can be distinguished owing to the different vibration energies between these two isotopes. Note that the h-BN phonon polariton signals at 179-193 meV present the homogenous intensity over the entire field of view due to the high delocalized nature of phonon polaritons (hundreds of nanometers). It also should be pointed out that during the sample transfer process, a twist between two h-BN isotopic nanosheets is introduced, for which we have experimentally measured (here ~10°) and compared with the DFPT calculation to evaluate the possible effect on the quantitative analysis (see details in supplemental figure 4 and Note 3). We find that although the in-plane phonons are sensitive to the phonon polaritons of h-BN [42], as shown in Figs. 1(c, d) and supplemental figure 4, the out-of-plane modes are not. Below we quantitatively analyze ZO phonon changes across the isotopic interface.

Fig. 2 shows the atomically resolved EELS of ZO phonons across the isotope interface. Fig. 2a is the HAADF image of the spectrum region and Fig.2b is the line profile of ZO phonons across the interface with the corresponding simulation data shown below (details in method). A dashed line highlights the isotope interface position. Two main spectral features, whose energy lies between 95-105 meV and 80-90 meV, originate from different momentum transfer, i.e., q near Γ (BZ center, low q) and q near K (BZ boundary, high q) by comparing the out-of-plane signals of DFPT results (supplemental figure 2). Thus, the two optical phonon modes are labeled as $ZO_{low\ q}$ and $ZO_{high\ q}$ respectively. Fig. 2c shows two typical eigenvectors of these two ZO modes respectively (modes selection shown in supplemental figure 5). In Fig. 2b, both $ZO_{low\ q}$ and $ZO_{high\ q}$ spectra exhibit a substantial variance of intensity with the atomic period, in good agreement with the DFPT results, indicating the experimental EELS results have achieved a sub-unit-cell scale (<0.34 nm) in probing the isotope signals.

Interestingly, across the interface the transitions of both these ZO phonons are not atomically sharp in Fig. 2d, and their transition widths are different, i.e., ~3.34 nm for $ZO_{low\ q}$ and ~1.66 nm for $ZO_{high\ q}$, both of which exceed a single atomic layer ~0.33 nm for h-BN (details of transition width fitting are shown in methods). In contrast, the ground state DFT calculations in supplemental figure 6 suggested that the transition should be atomically sharp, suggesting the involvement of effects beyond harmonic approximation and Born-Oppenheimer approximation harmonic phonon on the electronic ground state.

To explore the electron phonon coupling at the interface, we calculated the phonon-induced differential charge density at finite temperature. When considering vibration of optical modes in polar material h-BN, the positively charged B and negatively charged

N atoms are displaced from the equilibrium position with different amplitudes and opposite directions, thus inducing an additional dipole (vibration dipole) [43], as illustrated in Fig. 2e. At the interface of isotope heterostructure, the change of mass causes a discontinuity of vibration amplitude [44], thus further causes a gradient of vibration dipole as well as accumulation of bound charge at the interface (Fig. 2e). To evaluate this effect, we calculated the differential charge density induced by each ZO mode under 300 K (see supplementary note 1). The line profile of differential charge density induced by ZO modes at $q = \Gamma$, $q = M$ and $q = K$ are shown in supplemental figure 7, respectively. As can be seen, interfacial charge induced by ZO phonon at BZ center is higher than that at BZ boundary, indicating stronger interfacial electron-phonon coupling of $ZO_{low\ q}$. Such a phonon-induced differential charge density at the isotope interface is another isotope effect due to the electron-phonon interaction, which in turn affects the lattice vibration at the interface.

We further explore the changes in phonon energy between atomic layers near the interface. Fig. 3a shows a HAADF image of the h-$^{10}$BN/h-$^{11}$BN interface, wherein each atomic layer is labeled. In Fig. 3b, the phonon energy line profiles plotted from h-$^{10}$BN to h-$^{11}$BN show the phonon energy changes between the adjacent atomic layers. Specifically, from the "on-column" position to the "off-column" position, the measured phonon frequency is slightly increased. The quantitative analysis shown in Fig. 3c and 3d (see details in supplemental figure 8), where the averaged spectra from the bulk region with shadowed standard deviation error band, show that for both $ZO_{low\ q}$ and $ZO_{high\ q}$ the vibration energy goes up first and then goes down from one atomic layer to another. There is an energy difference ~0.68 meV of $ZO_{low\ q}$ between the on-column and off-column cases, which can be interpreted by the change in effective scattering angle [33]. However, the intra-atomic layer variation in $ZO_{high\ q}$ (Fig.3d) is not as significant as that in $ZO_{low\ q}$ (Fig.3c) because the signal of $ZO_{high\ q}$ comes from electrons scattered from the BZ boundary, where the phonon dispersion is flat so that the change of momentum transfer affects the energy little.

The phonons at the interface (black lines in Fig. 3c and 3d) are affected by both atomic scattering and mass change. Thus the vibration energy shows a combined and complicated manner. With the difference in the distance from the atomic interface, the dominant effect of scattering changes. As the probe moves from h-$^{10}$BN to h-$^{11}$BN, the cross-section of the electron is dominant when near the atomic layer, so the vibration energy rises first. Gradually approaching the h-$^{11}$BN layer, the influence of atomic mass change becomes dominant, thus the energy gradually decreases. The energy of the adjacent layers near the interface is also affected, consistent with results in Fig.2.

**Discussion**

For an isotope-purified h-$^{10}$BN/h-$^{11}$BN interface system, the lattice vibrations near the interface are probed with sub-unit-cell resolution. Meanwhile, the phonon energies of ZO$_{high\ q}$ and ZO$_{low\ q}$ are different, allowing us to reveal the momentum transfer-dependent behavior as well. We find that the ZO phonon modes at the interface are delocalized to a few nanometers, and the BZ center's phonons are more delocalized than those at the BZ boundary. We also find the electron-phonon coupling at the BZ center is stronger than that at BZ boundary. Moreover, the vibration energy of ZO phonons also varies from one atomic column to another, i.e., it reaches the maximum in the middle of them, which can be understood by the varying effective scattering angle during electron beam scattering. Such behavior also depends on the momentum transfer, that is, at the BZ boundary the flatter phonon dispersion makes the vibration energy less sensitive to the change of scattering angle in different positions. Across the interface from the h-$^{10}$BN layer to h-$^{11}$BN layer, the vibrational energy of two ZO phonons dramatically changes because both of the simultaneous changes in effective scattering angle and mass across the interface.

It is generally believed that the isotopes have little influence on electronic structures. A previous study reported that the optical band gap energies of h-BN depend on the isotopic composition [2], suggesting a strong nuclei effect on the electronic structures. Our study indicate that the electronic structure can be also affected by the isotopes but in a different manner, i.e., isotopes have different vibration amplitude and thus different vibration dipole, leading to the charge accumulation at their interface due to the discontinued van der Waals bonding, which in turn, delocalize the vibration behavior. More interestingly, such behavior is momentum-transfer dependent. Besides the phonon transport, the accumulated charge at the isotope interface likely also influences on the material electron activities.

On the other hand, it's well known that the isotope components govern the thermal conductivity of a material, and the well-accepted mechanism is that the mass disorder dampened the phonon transport which subsequently decreases the thermal conductivity [26, 27]. For an ideal interface, a native model would assume an atomically abrupt change of phonon energy which leads to the reduction of phonon transport channels due to reduced overlap of phonon DOS based on the diffuse mismatch model [45, 46]. However, the delocalization of lattice vibration at the isotope interface revealed in our work, is similar to that of the localized interface phonon modes for regular heterostructures such as c-BN/diamond [47] and Si/Ge [48], which can act as phonon bridges [49] to promote heat transport, suggesting isotope interface as a new degree of freedom for engineering of heat transport. On the other hand, in thermoelectrics, the highly desired suppression of delocalization phonons is generally realized by the introduction of structural defects and alloying-induced disorder. In this sense, our study suggests that the elimination of the isotope interface is also essential.

Moreover, considering the phonon group velocity also highly relies on the momentum transfer, the momentum transfer dependent delocalization of phonons at the isotope interface will affect heat transport in an even more complicated manner. These findings provide a new angle rather than a simple mass disorder to understand the isotopic effects on the physical properties in natural materials, and further shed light on tailoring material properties via proper engineering of isotopic interfaces.

## Method

**Crystal growth of isotopically tuned hBN and TEM sample preparation.** The enriched h-BN samples were synthesized using Fe flux method [43, 50]. High purity $^{10}$B (97.18 at %, 3M) or $^{11}$B (99.69 at %, 3M) and Fe (99.9%, Alfa Aesar) powders were mechanically mixed, and then loaded into an alumina crucible and placed at the center of a high-temperature single-zone tube furnace. The furnace was evacuated and then filled with $N_2$ (5 % $H_2$ mixed) and argon to atmospheric pressure. The furnace was heated to 1550 °C for a dwell time of 24-48 hours, ensuring precursors and Fe flux formed a complete solution. After that, a slow cooling to 1450 °C with a rate of 4 °C/hour is applied to obtain high-quality crystals. The furnace is then quenched (300 °C /hour) to room temperature. During the growth process, the $N_2$ (5 % $H_2$ mixed) and argon continuously flowed through the system at rates of 95 sccm and 5 sccm, respectively. The isotopically mixed h-BN crystals as required can be obtained by simply customizing the ratio of $^{10}B$ and $^{11}B$ powders.

The van der Waals heterostructures of h-$^{10}$BN/h-$^{11}$BN were prepared with polycarbonate (PC) thin film by a pick-up method. The h-BN nanosheets were first exfoliated on silicon substrates with a 300 nm-thick oxide layer. We used PC thin film to pick up the top h-$^{10}$BN nanosheet. The PC thin film with the h-$^{10}$BN nanosheet was then stamped onto the bottom h-$^{11}$BN nanosheet with accurate alignment control based on a homemade transfer stage and optical microscope. Then we dissolved the PC in trichloromethane for 12 h at RT. Finally, the as-transferred heterostructure was thermally annealed at 400 °C for 2 h under a high vacuum to further clean the surface. The stacked h-$^{10}$BN/h-$^{11}$BN nanosheet was then made into a TEM cross-section sample by Focused ion beam (FIB).

**EELS and imaging experiments.** The EELS experiments were carried out on a Nion U-HERMES200 electron microscope with a monochromator that operated at 30 kV and 60 kV to avoid damaging BN materials. For the sub-unit-cell resolution STEM-EELS experiments, we employed a convergence semi-angle $\alpha = 35$ mrad and a collection semi-angle $\beta = 25$ mrad, operated at 60kV. In this setting, the spatial resolution was better than 0.1 nm, while the energy resolution was ~8 meV. The beam current used for EELS was ~10 pA, and the acquisition times were 200-600 ms/pixel.

**EELS data processing.** All the acquired vibrational spectra were processed by using the custom-written MATLAB code and Gatan Microscopy Suite. More specifically, the EEL spectra were firstly aligned and then normalized to the intensity of the ZLP. Subsequently, the block-matching and 3D filtering (BM3D) algorithms were applied for removing the Gaussian noise. The background arising from both the tail of the ZLP and the non-characteristic phonon losses was fitted by employing the modified Pearson-VII function with two fitting windows and then subtracted in order to obtain the vibrational signal [47]. The Lucy-Richardson algorithm was then employed to ameliorate the broadening effect induced by the finite energy resolution, taking the elastic ZLP as the reference point for the spread function. The spectra were summed along the direction parallel to the interface for obtaining line-scan data with a good signal-to-noise ratio. In addition, we employed a multi-Gaussian peak fitting method to extract the polariton peaks from the composed signal.

**Interface transition width fitting.** The transition width of out-of-plane phonon modes is evaluated by a Logistic function, which is widely used to track the transition between two states. The formula of the employed fitting function is

$$y = \frac{a}{e^{\frac{(x-b)}{c}} + 1} + d$$

where $a, b, c, d$ are the fitting parameters. Among the four fitting parameters, $a$ and $d$ are used to normalize the dataset, $b$ is the center of two states, and $c$ is related to the conversion speed from one state to another. When the bias to the center is larger than $|\pm3c|$ then the concentration of one state is larger than 97.6%, thus we use $6c$ as the transition width.

## Acknowledgments

This work was supported by the National Key R&D Program of China (2021YFA1400500), the National Natural Science Foundation of China (52125307, 11974023, 52021006, 11974024, 92165101, 52025023, U1932153, 11974001), Guangdong Major Project of Basic and Applied Basic Research (2021B0301030002), the Strategic Priority Research Program of Chinese Academy of Sciences under Grant

No. XDB33000000, and the "2011 Program" from the Peking-Tsinghua-IOP Collaborative Innovation Center of Quantum Matter. We acknowledge the Electron Microscopy Laboratory at Peking University for using electron microscopes, and the High-performance Computing Platform of Peking University for providing computational resources.

## Data availability

The data that support the findings of this study are available from the corresponding author upon request.

## Code availability

Custom matlab codes that is used for data processing and DFT related post-processing are available from the corresponding author upon request.

## Author contributions

P.G., L.L., J.C., and E.G.W. conceived the idea; EELS experiments were performed by N.L. under the direction of P.G; growth of h-BN isotopes and the fabrication of h-BN heterostructure were performed by Y.F.L. under the direction of L.L; TEM samples were prepared by Z.T.L. and F.C.L.; DFT calculations were performed by R.C.S. and R.S.Q, under the direction of J.C. and E.G.W.; data processing and analysis were performed by N.L. and R.C.S., with assistant from R.S.Q , Y.H.L., X.D.G., and X.W.Z.; N.L. and R.C.S. wrote the manuscript with input from P.G., L.L., Y.F.L., K.H.L., and Y.J.; N.L., R.C.S. and Y.F.L. contributed equally to this work. E.G.W. supervised the project. All authors discussed the results at all stages and participated in the development of the manuscript.

## Competing interests

The authors declare no competing financial interests.

## Materials & Correspondence.

Supplementary information is available in the online version of the paper. Reprints and permissions information is available online at www.nature.com/reprints. Publisher's note: Springer Nature remains neutral with regard to jurisdictional claims in published maps and institutional affiliations. Correspondence and requests for materials should be addressed to J.C., L.L., E.G.W. or P.G.

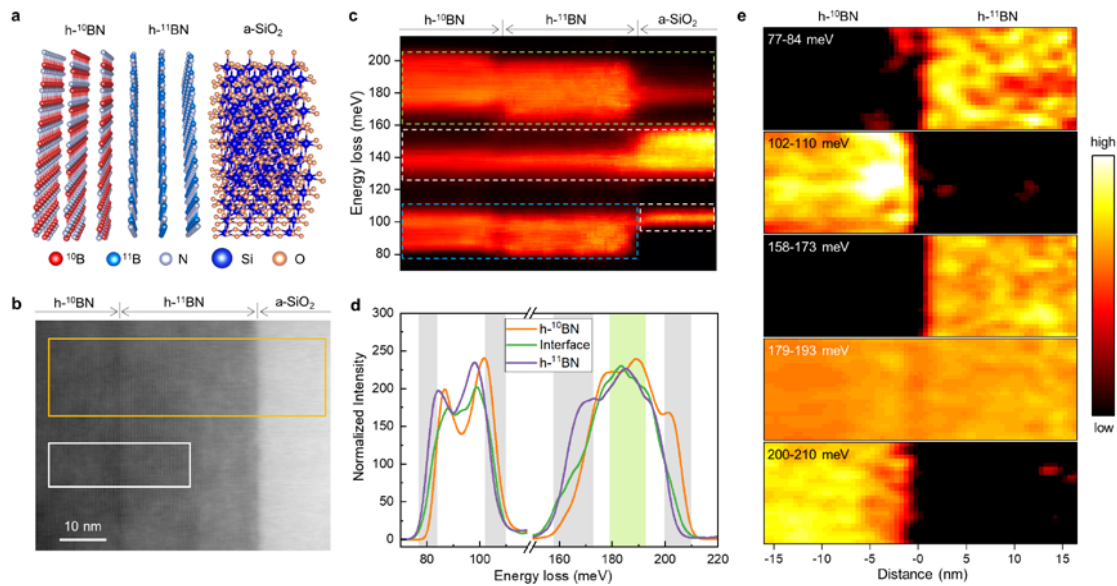

**Figure 1 Isotope identification of h-$^{10}$BN/h-$^{11}$BN interface.** **(a)** Schematic of the assembled stack of h-$^{10}$BN, h-$^{11}$BN and Si substrate with amorphous SiO$_2$ (a-SiO$_2$) layer on surface. **(b)** HAADF image of the h-$^{10}$BN/h-$^{11}$BN/a-SiO$_2$ cross-section. **(c)** STEM-EELS vibrational spectra were acquired at the region of the orange rectangle labeled in (b). The vibrational signals from the h-BN in-plane direction, a-SiO$_2$ and h-BN out-of-plane direction are labeled by green, white and blue dashed rectangles, separately. **(d)** STEM-EELS vibrational spectra. The orange, green, and purple solid lines are acquired from h-$^{11}$BN, interface, and h-$^{10}$BN, separately. The details of acquisitions and processing procedures are included in methods. **(e)** Energy-filtered imaging of h-$^{10}$BN/h-$^{11}$BN. The acquisition region is labeled by the white rectangle in (b), and the energy selection windows are shown as vertical stripes in (d) and labeled on the left-up corner of each map.

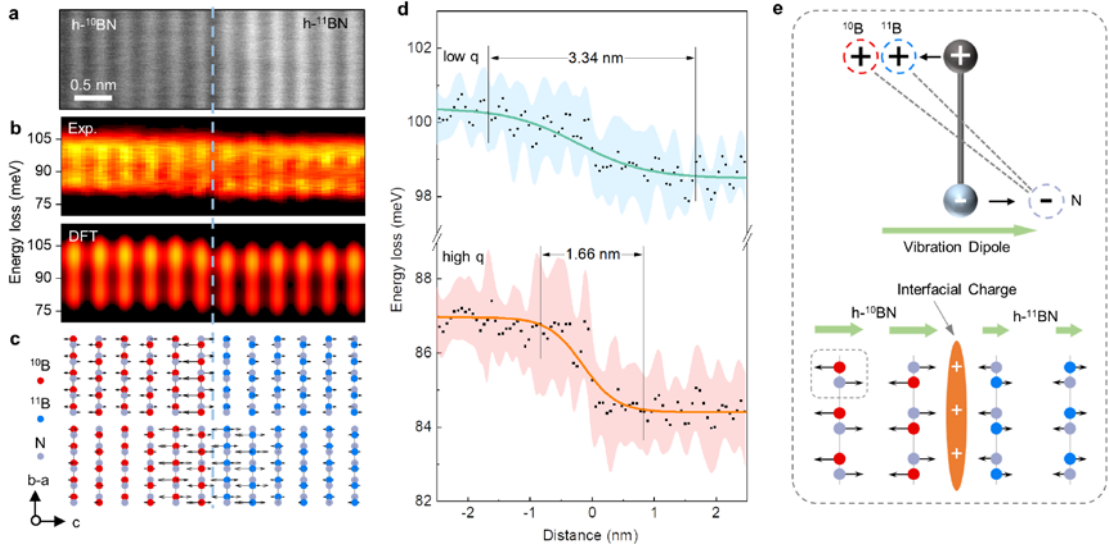

**Figure 2 Atomically quantitative analysis of ZO modes at h-$^{10}$BN/h-$^{11}$BN interface.** (**a**) HAADF image of h-$^{10}$BN/h-$^{11}$BN interface. (**b**) The corresponding experimental EELS obtained at the same region of (a), and the corresponding first principles calculation results with the same experimental parameters. See analysis details in methods. (**c**) Spatial distribution of typical phonon modes around the h-$^{10}$BN/h-$^{11}$BN interface. The arrows are eigenvectors extracted from DFT calculations corresponding to the Brillouin Zone center (q = Γ, ω = 102.468 meV) and Brillouin Zone boundary (q = K, ω = 71.201 meV & ω = 74.691 meV), denoting the two out-of-plane vibrations in (b). (**d**) Quantitative energy variation of the ZO modes. The black dots are fitted phonon energy of each spectrum, and the error bars are standard deviation shown as the orange and cyan shades. The orange and cyan solid lines are fitting of black dots by the Logistic function, presenting the transition width of $ZO_{low\ q}$ (Brillouin Zone center) and $ZO_{high\ q}$ ( Brillion Zone boundary) is 3.44 nm and 1.66 nm respectively. See analysis details in methods. (**e**) Schematic of phonon-induced vibration dipole and accumulated bound charge at interface caused by the discontinuity of atom displacement. Detailed spatial distribution of the charge density refer to supplemental figure 7.

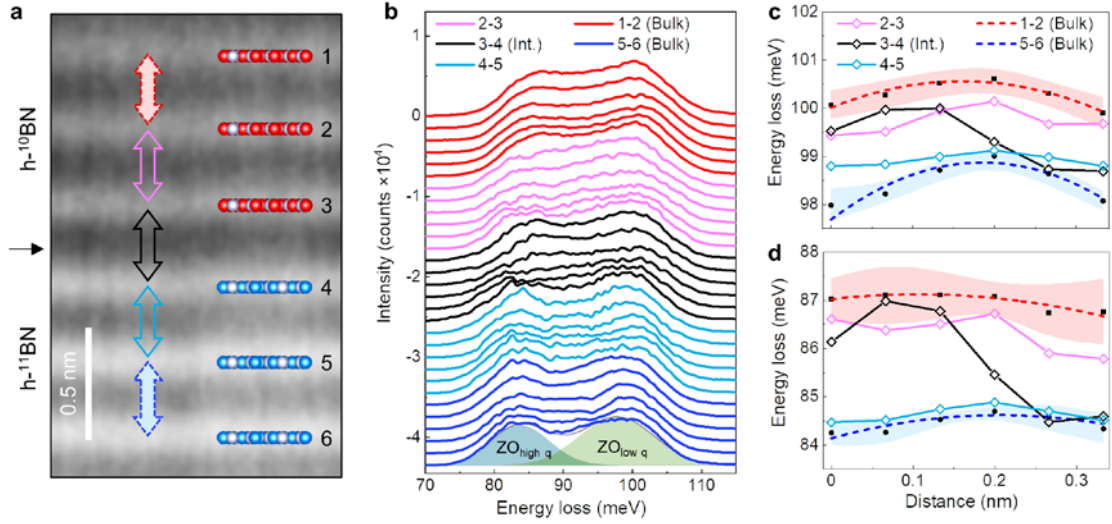

**Figure 3 Vibration energy change between atomic layers. (a)** HAADF image and the schematic of the structure of the h-$^{10}$BN/h-$^{11}$BN interface. Each atomic layer is labeled. The interface is between 3$^{rd}$ and 4$^{th}$ layer. **(b)** EEL spectra of ZO phonons at the bulk region of h-$^{10}$BN (red, averaged by multi layers away from the interface), interface (black, layer No. 3-4 ), adjacent layer to the interface (magenta and cyan, layers of No.2-3 and No.4-5), and h-$^{11}$BN (blue, averaged by multi layers away from the interface). No. "n-n+1" lable the spatra acquired between n and n+1 atomic layers. The bottom shows the fitted ZO$_{high\ q}$ and ZO$_{low\ q}$ modes by the multi-gaussian method. **(c, d)** The vibration energy change of (c) ZO$_{low\ q}$ and (d) ZO$_{high\ q}$ between the atomic layers. The starting position 0 corresponds to the No. n atomic layer, and the ending position correponds to the No. n+1 layer. The black dots are averaged phonon energy of multi layers in h-$^{10}$BN and h-$^{11}$BN, the red and blue dashed lines are fitted quadratic curves. The orange and cyan bands are the error bars calculated by standard deviation. Black (layer No. 3-4), magenta (layer No. 2-3) and cyan (layer No. 4-5) solid lines are data at the interface and two adjacent layers. The acquisition positions of each line are shown as colored arrows in (a).

# Supplementary information

**Note 1: First-principles calculations**

*Structure and phonon calculations.* Density functional theory (DFT) calculations of the structure and phonon properties of h-BN bulk and isotopic heterojunction were performed using Quantum ESPRESSO [1, 2] the Vanderbilt ultrasoft pseudopotentials [3]. The local density approximation (LDA) with the Perdew-Zunger parameterization [4] and the generalized gradient approximation with the Perdew-Burke-Ernzerhof [5] were considered for the exchange-correlation functional. The kinetic energy cut-off was 50 Rydbergs (Ry) for wavefunctions and 500 Ry for charge density and potential. A $9 \times 9 \times 3$ k-points mesh was used for bulk hBN, while a $6 \times 6 \times 1$ k-points mesh was used for the heterojunction. The heterojunction model contains 10 unit-cells of h-$^{10}$BN and 10 unit cells of h-$^{11}$BN connected along z direction (80 atoms in one hexagonal unit cell). The structure was optimized until the residual force was below $10^{-4}$ Ry per Bohr on every atom. The dynamic matrices and force constants were obtained using density functional perturbation theory (DFPT). The spatially resolved EELS cross-section calculation in Fig. 2b was performed following the same formula as literature [6].

*Phonon induced differential charge density calculation.* To calculate momentum- and mode-resolved phonon-induced differential charge density, we first interpolate the dynamic matrix of heterojunction on a $10 \times 10 \times 1$ q-mesh, and diagonalize it to get the eigenvector $\xi_{i,\alpha,\lambda,q}$, representing the eigenvector of $i$th atom, $\lambda$th mode at $q$ point in Cartesian coordinate direction $\alpha$. Then the atomic displacements are generated based on $d_{i,\alpha,\lambda,q} = \xi_{i,\alpha,\lambda,q}/\sqrt{m_i}$, where $d_{i,\alpha,\lambda,q}$ is the displacement and $m_i$ is the mass of $i$th atom. The charge density of each displaced configuration is then calculated using DFT with the same parameter as described before. The thermal excitation coefficients under finite temperature are calculated by molecular dynamics using the LAMMPS package [7] with the Tersoff interatomic potential parametrized in ref. [8]. The displacement of $i$th atom $x_{i,\alpha}$ can be written as $x_{i,\alpha}(t) = \sum_{\lambda,q} \xi_{i,\alpha,\lambda,q} \cdot \frac{1}{\sqrt{m_i}} \cdot e^{-i\omega_{\lambda,q}t} \cdot C_{\lambda,q}$, where $\omega_{\lambda,q}$ is the frequency of $\lambda$ th mode at $q$ point, $C_{\lambda,q}$ is the thermal excitation coefficient. Using the fact $\sum_{i,\alpha} \xi_{i,\alpha,\lambda,q}^* \cdot \xi_{i,\alpha,\lambda,q} = 1$, we can get $C_{\lambda,q} = \sum_{i,\alpha} \xi_{i,\alpha,\lambda,q}^* \cdot \sqrt{m_i} \cdot X_{i,\alpha,\lambda,q}$, where $X_{i,\alpha,\lambda,q}$ is the $\lambda,q$ component of the Fourier transform of $x_{i,\alpha}(t)$. The charge density of low-q modes and high-q modes are averaged with weight $C_{\lambda,q}$ respectively, where low-q modes are defined as its distance to Γ point is smaller than the half of the side length of Brillouin zone, and the other modes are defined as high-q modes. The differential charge density is then obtained by the difference between the averaged charge density under phonon perturbation and the charge density of unperturbed structure. For each configuration, the differential charge density of perturbation along the positive and negative, i.e., +z direction and -z direction, is averaged. The interfacial charge is calculated as the integral of the differential charge density over one adjacent unit cell (~0.34 nm) of the interface each side. The line profile of differential charge density is convoluted with a gaussian factor considering the limited beam size.

**Note 2: Influence of the SiO$_2$ signals on h-BN isotopic phonon analysis**

As shown in the Supplemental Figure 3b below, the peak at ~130 meV is the phonon polaritons of SiO$_2$, which can be detected in vacuum under aloof mode [9]. The energy window of this signal has no overlap with h-BN, i.e., 160-200 meV for in-plane and 80-100 meV for out-of-plane. Thus, it would not affect the analysis of h-BN phonons. The peak at ~100 meV is predominantly excited by impact scattering [10], which decays quickly with no energy variation [9]. Therefore, the very short-range interaction of this signal makes little influence to our analysis on the out-of-plane phonons of h-BN.

**Note 3: Influences of the twist angle on the vibration signals**

As shown in the Supplemental Figure 4 below, we built h-BN cross-sectional structures with different crystal axis directions, including [1 0 0], [10 1 0], [5 1 0], [10 3 0], [5 2 0] and [2 1 0]. The corresponding tilt angles are 0°, 5.74°, 11.54°, 17.46°, 23.58° and 30°, respectively. These typical tilt angles can basically reflect the trend of the influence on the phonon signals variation in EELS. The EELS vibrational spectrum calculated by DFPT is shown in Fig. S4b, where colored arrows highlight the dominant vibration modes in each spectrum. In order to study the influence of the zone-axis deflection on vibrational spectrum, we quantitatively extracted peak position and signal intensity of the main vibration modes. Fig. S4c shows the change of the energy of each mode as the tilt angle increases, and Fig. S4d shows the change of the intensity of each mode with the tilt angle increases.

When the tilt angle reaches the maximum 30°, the energy of TO mode has a blue shift up to ~10 meV, and its intensity changes by more than 2 times (as seen in the middle panels of Fig. S4c, d). However, when the tilt angle is small, the energy shift is not significant. For example, with 5° tilt, the energy shift of TO mode is less than 0.2 meV, which is even small than the typical precision of EELS thus can be neglected during the quantitative analysis for the in-plane vibrational signals, while for the ZO modes along out-of-plane direction, the energy shift is smaller than 1.1 meV. So the tilt angle between the heterostructure needs to be controlled to a small extent in the experiments. In addition, we find that the energy changes of in-plane modes are non-linear, while the out-of-plane modes show linear relationship to the tilt angle.

In fact, in our experiments the sample have always been rotated to make the electron beam incident at the middle angle of two isotope flakes. Under this circumstance, the tilt angle between sample and electron beam is halved. Therefore, the two parts have the same tilt angle thus do not affect the comparison of signals between the two parts. For example, for data presented in the manuscript, the angle between the h-$^{10}$BN/h-$^{11}$BN isotopes is experimentally measured to be ~10° by rotating the h-$^{10}$BN h-$^{11}$BN part successively to zone axis with the assistance of convergent-beam electron diffraction. The sample has been rotated to the middle of h-$^{10}$BN/h-$^{11}$BN layer, i.e., the angles between the electron beam and the h-$^{10}$BN /h-$^{11}$BN layer are both ~5°. In this case, we estimate the angle deviation between these two flakes should be smaller than ~5°, for which the angle effect can be completely negligible compared the instrumental precision.

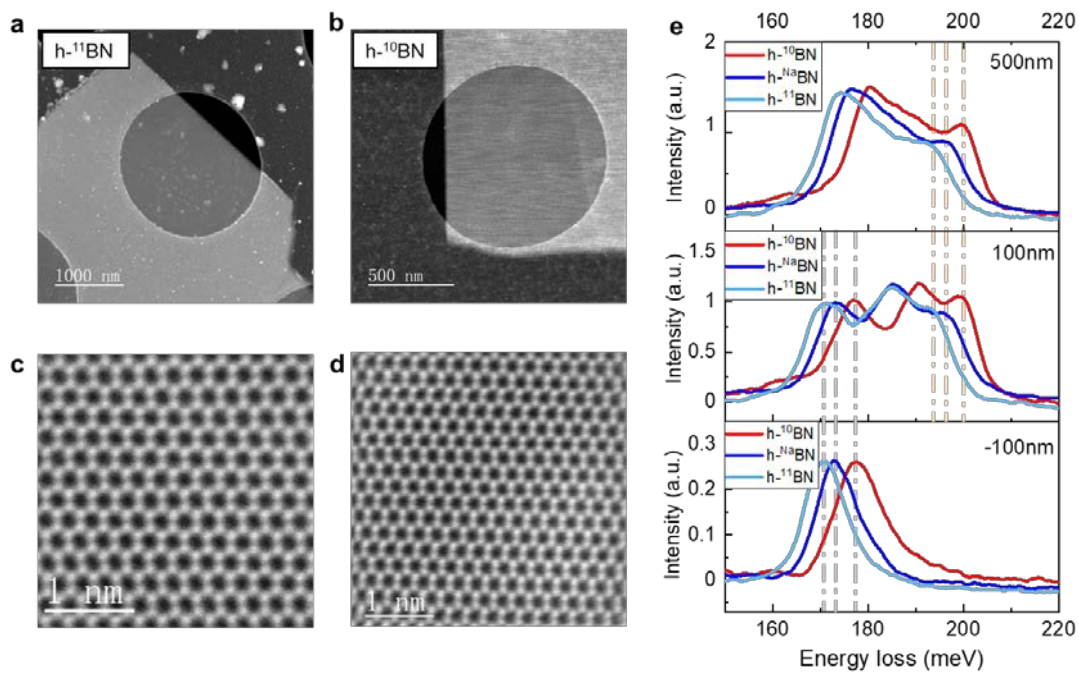

**Supplemental figure 1 | Isotope enriched hBN flakes. (a, b)** Low magnification HAADF image of h-$^{11}$BN and h-$^{10}$BN crystals. **(c, d)** Atomically resolved HAADF image of multilayer h-$^{11}$BN and h-$^{10}$BN, showing high quality. The image has been denoised using Gaussian blur. **(e)** Vibrational STEM-EELS spectra of h-$^{10}$BN, h-$^{Na}$BN and h-$^{11}$BN at different positions, presenting the good isotopic separation ability of EELS. Negative distance corresponds to the aloof mode acquisition. The dashed vertical grey lines denoting the signals of phonon polariton, orange lines denoting the signals of LO phonon modes of h-BN.

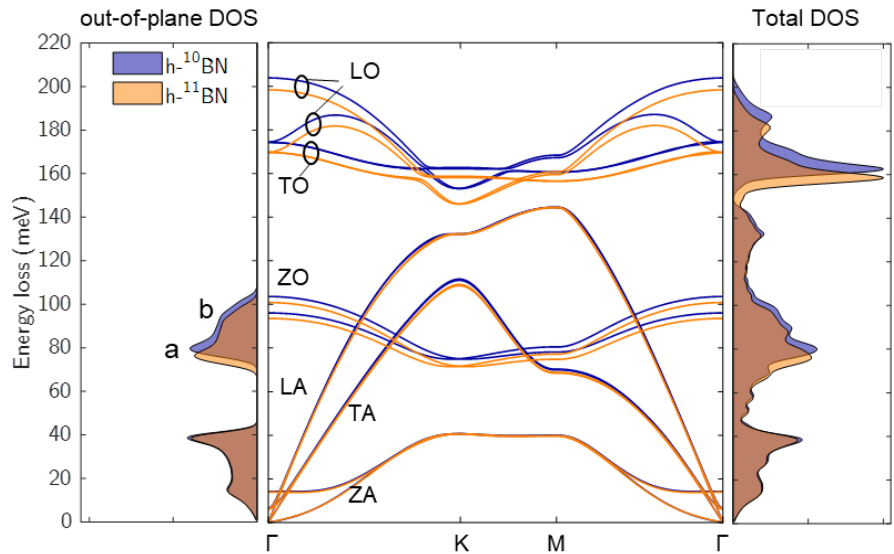

**Supplemental figure 2** | DFT calculated phonon dispersion of h-$^{10}$BN and h-$^{11}$BN isotopes, and corresponding phonon density of states (DOS). The left part is the DOS of out-of-plane component and the right part is the total DOS of all directions. At the range of 80-100 meV there are two dominate peaks a and b, which are the out-of-plane optical phonons (ZO). Peaks a and b are originated from phonons with different q, thus we labeled them as ZO$_{high\ q}$ (at BZ bounary) and ZO$_{low\ q}$ (at BZ center) in the manuscript.

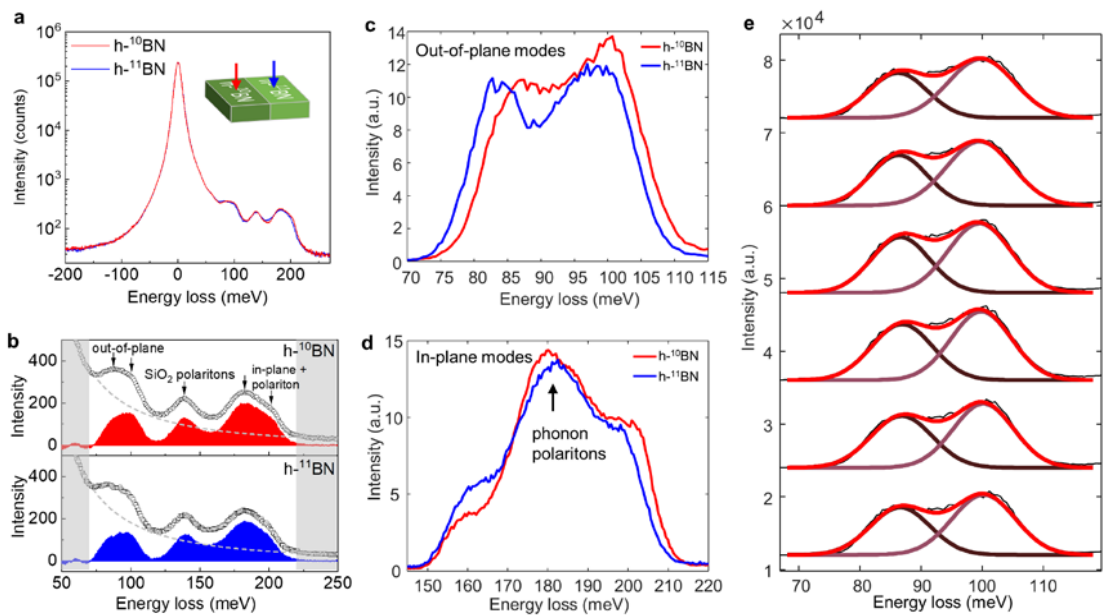

**Supplemental figure 3 | EELS processing and the fitting of ZO modes. (a)** Original EELS spectra acquired at h-$^{10}$BN region and h-$^{11}$BN region in log scale. **(b)** The dots are summed raw EELS data, dashed lines are the fitted background using the modified Pearson-VII function (fitting windows are 55-70 meV and 220-250 meV), the filled red and blue curves are the background-subtracted signals. **(c, d)** The background-subtracted signals of out-of-plane and in-plane modes de-convoluted using Lucy-Richardson algorithm. The in-plane modes are significantly influenced by the phonon polaritons of h-BN, of which the signal energy is sensitive to the edge distance and sample geometry. **(e)** Multi-gaussian fitting of the out-of-plane signals. These spectra are acquired at different positions between two atomic layers.

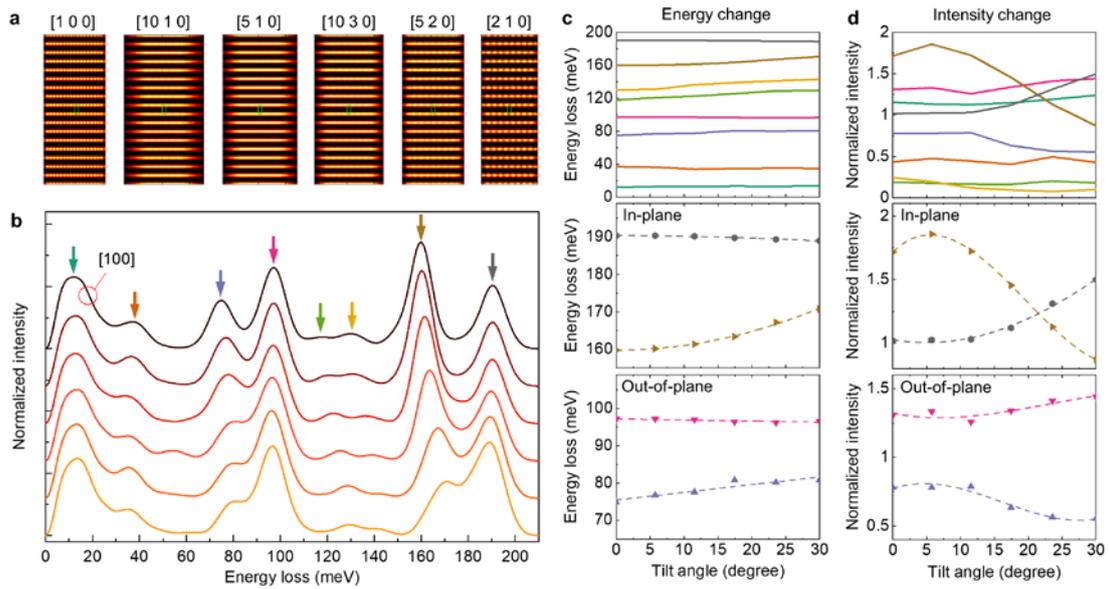

**Supplemental figure 4 | Influence of sample tilt on the vibrational spectrum measurement.** (**a**) Schematic diagram of the cross-section h-BN at different tilt angles. (**b**) EELS vibrational spectra at different tilt angles calculated by DFPT. (**c**) The relationship between tilt angle and the energy of different vibrational modes. The in-plane modes and out-of-plane modes are highlighted in the middle and the bottom panels, respectively. With tilt angle at 30°, the maximum energy change is ~10 meV for the in-plane modes, and ~5meV for the out-of-plane modes. In addition, the energy changes of in-plane modes are non-linear, while the out-of-plane modes show linear relationship to the tilt angle. (**d**) The relationship between tilt angle and the intensity of different vibrational modes. The in-plane modes and out-of-plane modes are highlighted in the middle and the bottom panels, respectively.

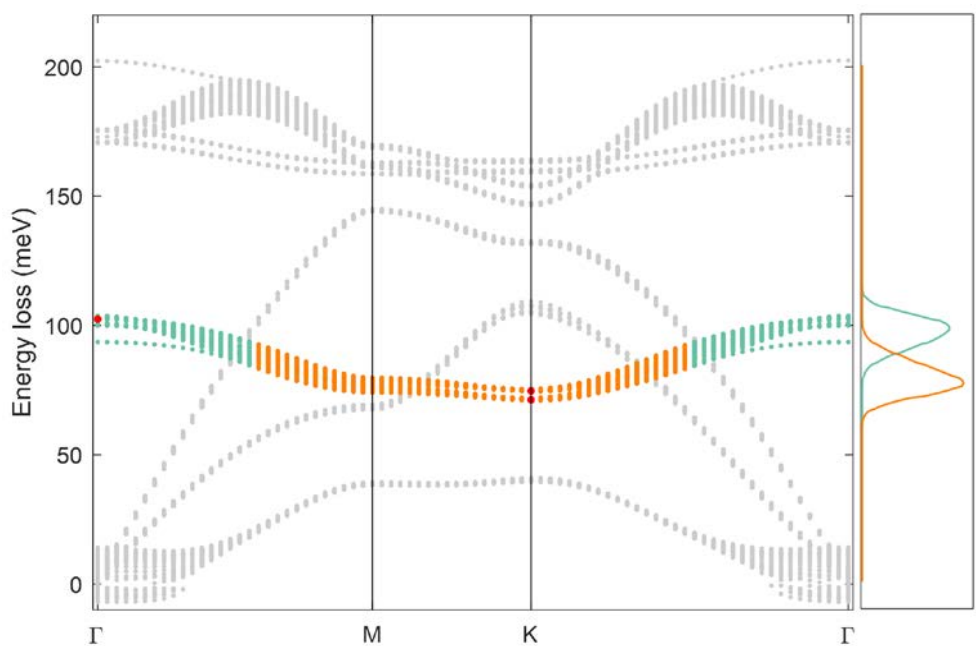

**Supplemental figure 5 | Vibration modes of h-BN calculated by DFPT.** In the left panel, each dot represents one phonon mode, while the modes highlighted by red color correspond to the typical Brillouin Zone center (q = Γ, ω = 102.468 meV) and Brillouin Zone boundary (q = K, ω = 71.201 meV & ω = 74.691 meV) out-of-plane optical phonon modes persented in Fig. 2c. Low q modes are labeled by green and high q modes are labeled by orange. The right panel shows the out-of-plane phonon DOS corresponding to low q and high q modes respectively.

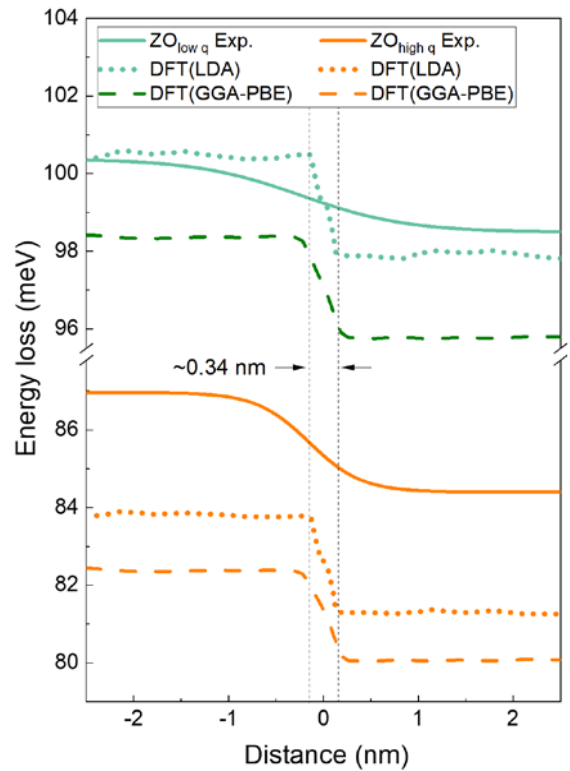

**Supplemental figure 6 | Comparison of DFPT Calculations and the experimental ZO phonon tansitions.** The dotted cyan and orange lines are the energy change of ZO modes around the interface calculated by DFT with LDA functional, while the dashed ones are calculated by DFT with GGA-PBE functional. The solid lines are fitted from experimental results. Both DFT calculations show a very sharp transition at the interface, whose transition length is ~0.34 nm (one atomic layer), suggesting phonon transition at the isotopic interface of h-BN involves effects beyond harmonic approximation and Born-Oppenheimer approximation. It is worth mentioning that the phonon energy difference between the two calculations and experimental results is about 5 meV. Such a deviation is likely to be mainly caused by the limitation of accuracy of the exchange-correlation functional used. Improvement of DFT methods are needed to reach quantitative agreement with experiment (within meV) for phonon dispersion, but the two calculations suggest the width of the phonon transition layer is not affected by the choice of the exchange correlation functional.

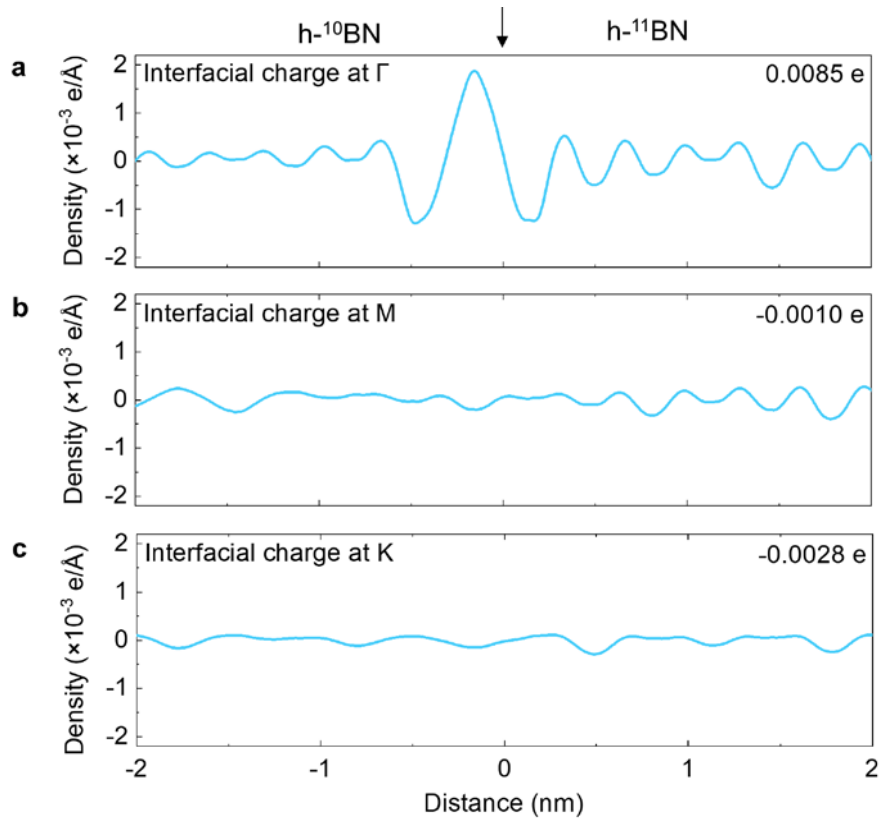

**Supplemental figure 7 | Phonon induced differential charge density at h-$^{10}$BN/h-$^{11}$BN interface.** The line profile of differential charge density induced by ZO phonon at **(a)** Γ, **(b)** M and **(c)** K point along the [0001] direction. Numbers at right-top corner of each panel show the total interfacial charge in the adjacent unit cell of the interface each side. As can be seen, interfacial charge induced by ZO phonon at Γ point is much higher than that at M/K point, indicating stronger interfacial electron-phonon coupling of ZO$_{low\ q}$.

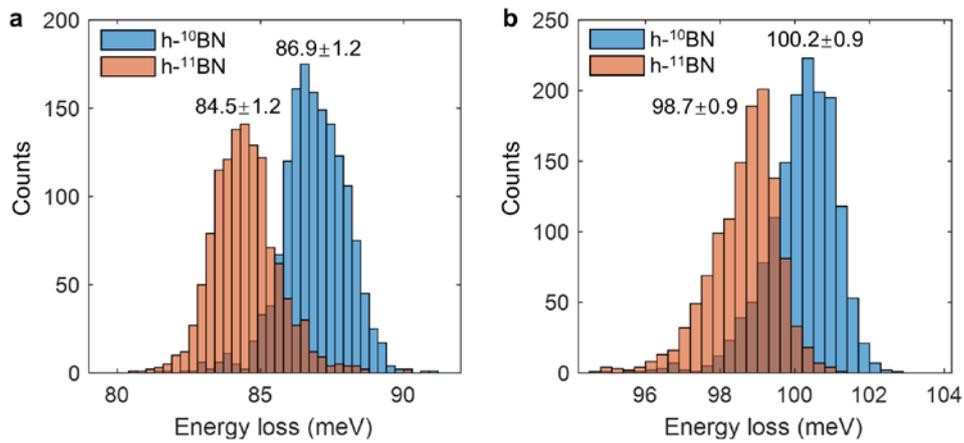

**Supplemental figure 8 | Fitted energy distributions of $ZO_{high\ q}$ and $ZO_{low\ q}$.** Histogram of fitted peak positions of **(a)** $ZO_{high\ q}$ and **(b)** $ZO_{low\ q}$ from 1500 acquisitions in h-$^{10}$BN/h-$^{11}$BN heterostructure, demonstrating clear separation of the isotopes.